\documentclass{aastex701}

\usepackage{lipsum}
\usepackage{adjustbox}
\usepackage{caption}
\usepackage{url}
\usepackage{hyperref}
\usepackage{amsmath}
\usepackage{float}
\numberwithin{equation}{section}
\usepackage{amssymb}
\usepackage{rotating}

\hypersetup{
    colorlinks=true,
    linkcolor=blue,
    citecolor=blue,
    urlcolor=blue
}
\begin{document}
\title{Solar Vortex Detection With Velocity Field Normalisation: Eliminating False Positives}

\author[0009-0000-4129-1324]{Lauren McClure}
\affiliation{Plasma Dynamics Group, 
School of Electrical and Electronic Engineering, 
The University of Sheffield, Sir Frederick Mappin Building, Mappin Street, Sheffield, S1 3JD}
\email{lmcclure1@sheffield.ac.uk}

\author[0000-0001-5414-0197]{Suzana Silva}
\affiliation{Plasma Dynamics Group, 
School of Electrical and Electronic Engineering, 
The University of Sheffield, Sir Frederick Mappin Building, Mappin Street, Sheffield, S1 3JD}
\email{suzana.silva@sheffield.ac.uk}

\author[0000-0002-9546-2368]{Gary Verth}
\affiliation{Plasma Dynamics Group, 
School of Mathematical and Physical Sciences, 
The University of Sheffield, Hicks Building, Hounsfield Road, Sheffield, S3 7RH}
\email{g.verth@sheffield.ac.uk}

\author[0000-0002-3066-7653]{Istvan Ballai}
\affiliation{Plasma Dynamics Group, 
School of Mathematical and Physical Sciences, 
The University of Sheffield, Hicks Building, Hounsfield Road, Sheffield, S3 7RH}
\email{i.ballai@sheffield.ac.uk}
 
\author[0000-0002-0893-7346]{Viktor Fedun}
\affiliation{Plasma Dynamics Group, 
School of Electrical and Electronic Engineering, 
The University of Sheffield, Sir Frederick Mappin Building, Mappin Street, Sheffield, S1 3JD}
\email{v.fedun@sheffield.ac.uk}

\begin{abstract}
Small-scale vortices in the solar photosphere play a central role in transporting mass, energy, and momentum into the upper solar atmosphere, yet reliably detecting these structures remains rather challenging. We address this problem by introducing a simple preprocessing step that normalises the velocity field by its magnitude. Our method preserves flow topology while suppressing shear-induced artefacts that lead to spurious detections in non-uniform, high-rotation environments. For validation, we apply this approach to high-resolution Bifrost simulations and evaluate vortex detection using four commonly employed methods: IVD, the $\lambda_2$-criterion, the Q-criterion, and the $\Gamma$ method. We assess which structures exhibit physically consistent rotation by using the $d$-criterion to automatically detect rotational plasma-flow features, which we use as an approximate ground truth. We find that, in the unnormalised field, a substantial fraction of detections made by the first three methods are false positive detections. Normalisation removes most of these. The $\Gamma$ method detects true vortices but misses a large number of vortical flows. The normalisation step yields better-defined and more realistic vortex boundaries. As the $\Gamma$ method underpins most observational analyses, current studies likely capture only a subset of vortical flows. By comparison, the other three methods detect four to five times more vortices after normalisation, suggesting that the true photospheric vortex coverage may be underestimated by a similar factor. Overall, this physically motivated preprocessing step enhances the accuracy and physical realism of vortex detection and offers a practical enhancement for analysing vortical flows in turbulent flows.

\end{abstract}


\section{Introduction}
Vortex tubes are ubiquitous phenomena in the solar atmosphere, where they serve as important channels for the transport of energy, mass, and momentum from the photosphere through the chromosphere and into the corona \citep{Tziotzou_2018}. This behaviour appears consistently across multiple magnetohydrodynamic (MHD) simulations. In essence, vortices are swirling columns of plasma, typically associated with upward flows in the chromosphere and corona, but predominantly linked to downward flows in the photosphere \citep{Bonet_2008, Wedemeyer_2012}. By definition, from a two-dimensional identification perspective, they have high local vorticity, are bounded, persistent in time, and exhibit circular (or quasi-circular) motion. Alongside these characteristics we have a general definition for solar vortex flows proposed by \citet{Tziotziou_2023}: “A vortex is the collective motion associated with the azimuthal component ($\varphi$) of a vector field (e.g. true velocity or magnetic field) or its observational counterparts (such as trackable motions of features associated with radiation intensity or magnetic field) about a common centre or axis.” Nevertheless, despite these recognised traits, there is still no universally accepted mathematical definition of a vortex. This has led to the development of numerous automated vortex identification methods \citep{Cuissa_2022}. Since the introduction of the Okubo–Weiss criterion, which evaluates whether strain or rotation dominates within a local flow region, a wide range of additional metrics have been proposed \citep{Okubo_1970, Weiss_1991}. The Q-criterion subsequently extended this approach by comparing the magnitude of rotation to that of strain, and can be generalised to three dimensions by evaluating velocity gradients through their normal components rather than eigenvalues \citep{Hunt_1988}. Both the Q-criterion and the Okubo-Weiss criterion are classically regarded as effective tools for identifying regions within a flow that exhibit vortex-like behaviour.

More recent methods have focused on locating a defined centre of a vortex within a flow field \citep{Silva_2018, Graftieaux_2001}, with some approaches also specifying the surrounding region in which the vortical behaviour is confined. A straightforward approach to identifying a vortex centre is to search for local maxima in the vorticity magnitude \citep{Strawn_1999}. While this method directly captures the ``high local vorticity” characteristic, it also risks misclassifying strong shear flows as vortical structures.

One widely used approach that identifies both the region of a vortex and its centre is the $\Gamma$ method \citep{Graftieaux_2001}. In this method, the vortex centre and boundary are defined based on the flow topology within a restricted domain by detecting regions that exhibit solid-body rotation around a central point. The LAVD (Lagrangian Averaged Vorticity Deviation) and IVD (Instantaneous Vorticity Deviation) methods (in Lagrangian and Eulerian forms, respectively), introduced by \citet{Haller_2016}, determine vortex centres by evaluating the deviation of vorticity within a domain from the mean vorticity, and locating the local minima and maxima of the resulting field. The corresponding vortex boundaries are obtained by identifying the outermost bounded isocontour that encloses these extrema, with the contours further filtered according to their convexity.

A wide variety of additional vortex detection methods have been developed, most of which emphasise one or two of the recognised vortex characteristics \citep{Cuissa_2022, Yuan_2023, ZHOU_1999}. In this paper, we introduce a modification to vortex detection approaches and apply our methodology to simulated vortex flows in the solar photosphere. Our focus is on finding a robust vortex detection methodology that remains reliable despite the asymmetry of rotational flows. By concentrating exclusively on the topology of the flow while disregarding its magnitude, we demonstrate a means of achieving more accurate vortex identifications, with potential applications in both terrestrial hydrodynamics and space-based magnetohydrodynamics. We illustrate the method using small-scale vortices in the solar photosphere, where asymmetric vortex flows are prevalent \citep{Moll_2011, Aljohani_2022}. 

\section{Methodology}
\subsection{Bifrost Simulation Data}
All results presented in this paper are obtained using numerical simulations performed with the radiative magnetohydrodynamical code Bifrost \citep{Gudiksen_2011}. Designed to reproduce the velocity and magnetic fields of the upper solar atmosphere with high accuracy, Bifrost enables detailed investigation of photospheric kinetic vortices. Because the photosphere is highly turbulent and dominated by numerous shear-like flows, any vortex detection method must be capable of operating reliably in turbulent conditions and of distinguishing effectively between shear and vortical motions.

The specific simulation used in this study is publicly available through the Hinode Science Data Centre (SDC) as part of the IRIS project \citep{Carlsson_2016}. The dataset, named ch024031\textunderscore by200bz005, spans a horizontal domain of 24 Mm $\times$ 24 Mm and a non-uniform vertical extent of 17 Mm, beginning 2.5 Mm below the solar surface. For this analysis, we focus on a photospheric layer located 0.44 Mm above the solar surface. At a spatial resolution of 31km, the domain for applying the two-dimensional vortex detection methods corresponds to 768 $\times$ 768 grid points.

\begin{figure*}[t!]
  \centering
  \captionsetup{font=small,skip=2pt}
  \begin{adjustbox}{max width=\textwidth, max totalheight=0.43\textheight}
    \includegraphics[width = \linewidth]{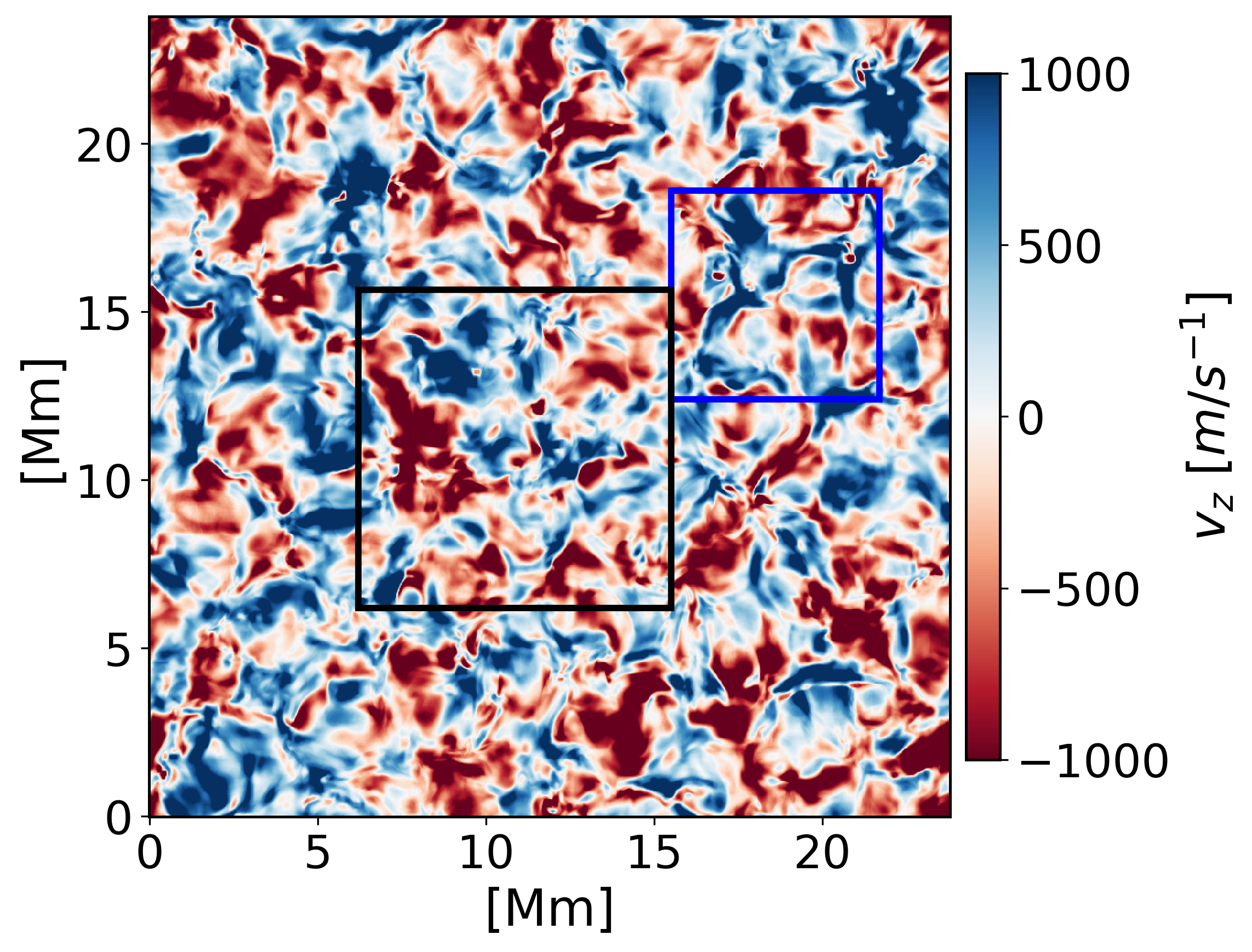}
  \end{adjustbox}
  \caption{A view of the vertical component of velocity, $v_z$, over the numerical domain over which we conduct our analyses. The blue and black box represent the subdomain shown in Appendix ( the blue blox is shown in Figure (\ref{fig:6}) and the black box represents the subdomain shown in Figures (\ref{fig:7}--\ref{fig:10})).}
  \label{vz_domain}
\end{figure*}

\subsection{Normalisation of the Velocity Field}
The modification we propose involves normalising the velocity field within the fundamental equations of each vortex detection method. This is achieved by dividing the $x$- and $y$-components of the velocity field by the velocity magnitude. The rationale is that by focusing exclusively on the flow topology at a given time, the vortex region can be identified with greater accuracy.

We can apply this modification in the two-dimensional and three-dimensional perspectives. In both cases, we divide the velocity vector, $\boldsymbol{U}$, at each grid point by the magnitude of velocity there, $U$. We define the magnitude of velocity using the Euclidean norm of the velocity vector. At each grid point, the velocity field will be normalised as  
\begin{equation}
    \boldsymbol{\tilde{U}} = \boldsymbol{U}/U.
\end{equation}
We then use the $\tilde{\boldsymbol{U}}$ as our velocity field when applying the necessary criteria for vortex identification. 

\subsection{Vortex Detection Methods}
In our study, we test the modification described above on four main vortex detection methods: the Q-criterion \citep{Hunt_1988}, $\lambda_2$-criterion \citep{Jeong_Hussain_1995}, IVD \citep{Haller_2016, Silva_2020} and $\Gamma$ method \citep{Graftieaux_2001}. For each of these methods, we will calculate the boundary and centre of a vortex using the normalised velocity field, $\tilde{\boldsymbol{U}}$, as well as with the actual velocity field. We choose these four identification methods as they are tried and tested methods that are commonly used in solar physics in the identification of small-scale vortices from particle image velocimetry measurements, see e.g. \citep{Tziotziou_2023}. 

\subsubsection{The Q-criterion}
The first vortex-detection method we consider is the Q-criterion, which compares local rotation with shear by decomposing the velocity-gradient tensor into its antisymmetric (rotation-rate tensor, $\boldsymbol{\Omega}$) and symmetric (strain-rate tensor, $\boldsymbol{S}$) parts \citep{Hunt_1988}. Defining
\begin{equation}
Q = \tfrac{1}{2}\bigl(\lVert \boldsymbol{\Omega} \rVert^{2} - \lVert \mathbf{S} \rVert^{2}\bigr).
\end{equation}
 Regions with $Q>0$ are rotation-dominated and are candidates for vortices. To minimise false positives, we restrict our attention to the upper tail of $Q$. Specifically, we retain as detections only those areas in the domain above the 90th percentile of the distribution of positive $Q$ values (i.e. the top 10$\%$ of $Q>0$).

\subsubsection{The $\lambda_2$-criterion}
The $\lambda_2$-criterion, introduced by \citet{Jeong_Hussain_1995}, identifies vortices as regions where the pressure field exhibits a local minimum in a plane perpendicular to the vortex axis. This is motivated by the fact that swirling motion creates a low-pressure core, whereas shear does not. Starting from the incompressible Navier–Stokes momentum equation, taking its gradient, and decomposing the velocity-gradient tensor into its symmetric and antisymmetric parts, $\boldsymbol{S}$ and $\boldsymbol{\Omega}$, yields
\begin{equation}
    \frac{d\boldsymbol{S}}{dt} - \nu \Delta \boldsymbol{S} + \boldsymbol{\Omega}^{2} + \boldsymbol{S}^{2}
    = -\frac{1}{\rho}\,\nabla(\nabla p),
    \label{NS}
\end{equation}
where $\rho$ and $\nu$ denote density and viscosity, respectively, and $\nabla(\nabla p)$ is the Hessian of the pressure field. This relates the local curvature of the pressure to the tensors $\boldsymbol{S}$ and $\boldsymbol{\Omega}$, allowing pressure minima and therefore vortex cores to be identified through the eigenvalues of $\boldsymbol{\Omega}^{2} + \boldsymbol{S}^{2}$.

Under the usual assumption of incompressibility, this relation shows that local extrema of pressure can be inferred from the eigenvalues of the symmetric tensor $\boldsymbol{\Omega}^{2} + \boldsymbol{S}^{2}$, allowing vortex cores to be identified based on their pressure signature rather than on vorticity magnitude alone.
 The local pressure minimum is only successful in its identification of a vortex when neglecting unsteady and viscous effects \citep{CHAKRABORTY_2005}, hence we look for pressure minima using
\begin{equation}
    \boldsymbol{\Omega}^2 + \boldsymbol{S}^2 = -\frac{1}{\rho}\nabla(\nabla p).
\end{equation}

The second-order tensors $\boldsymbol{S}$ and $\boldsymbol{\Omega}$ are defined from the velocity-gradient tensor, and their squares are taken using standard tensor (matrix) multiplication. Both $\boldsymbol{S}^2$ and $\boldsymbol{\Omega}^2$ are symmetric, and therefore their sum,
\[
\boldsymbol{S}^2 + \boldsymbol{\Omega}^2,
\]
is also symmetric. As a consequence, all three eigenvalues $\lambda_1 \geq \lambda_2 \geq \lambda_3$ of this tensor are real. The Q-criterion can be written in terms of these eigenvalues as
\begin{equation}
    Q = -\tfrac{1}{2}\mathrm{tr}(\boldsymbol{S}^2 + \boldsymbol{\Omega}^2) = -\tfrac{1}{2}(\lambda_1 + \lambda_2 + \lambda_3).
\end{equation}

Following \citet{Jeong_Hussain_1995}, vortex cores correspond to regions of locally low gas pressure, which in this formulation occur when two of the three eigenvalues of $\boldsymbol{S}^2 + \boldsymbol{\Omega}^2$ are negative. Since the eigenvalues are ordered, this requirement is equivalent to demanding
\[
\lambda_2 < 0.
\]
Thus, the $\lambda_2$-criterion identifies vortices as connected regions where the second largest eigenvalue of $\boldsymbol{S}^2 + \boldsymbol{\Omega}^2$ is negative.

Here we identify vortex centres as local minima of the $\lambda_2$ field that fall within the most-negative 25\% of values across the domain. We compute a global threshold $\lambda_2^\ast$ as the 25th percentile of the negative tail ($\lambda_2<0$), then build a binary mask $\{\lambda_2 \le \lambda_2^\ast\}$, remove regions smaller than 4 pixels, and extract the boundaries of the remaining regions. These boundaries are the perimeters of the thresholded set and, therefore, approximate the $\lambda_2=\lambda_2^\ast$ isocontours. This global thresholding on the negative tail reduces false positives while providing a consistent boundary estimate.

\subsubsection{The Instantaneous Vorticity Deviation (IVD)}
In 2D, the IVD field is defined as \citep{Haller_2016} 
\begin{equation}
    IVD(\boldsymbol{x},t) = |\omega_z(\boldsymbol{x},t) - \bar{\omega}_z(t)|, 
\end{equation}
which compares the $z$-component of vorticity, $\omega_z$, to its average value at each coordinate . This expression essentially looks for deviations away from the average vorticity that could be attributed to a vortex centre. Hence, the centre and boundary are then defined based upon local maxima of this field, where boundaries are the outermost bounded isocontour of the IVD field surrounding the maxima. This is conditioned by the convexity deficiency variable,
\begin{equation} \label{convex-deficiency}
    \frac{A(\mathcal{H})- A(\mathcal{C})}{A(\mathcal{C})},
\end{equation}
where $A(\mathcal{C})$ denotes the area within the detected boundary and $A(\mathcal{H})$ denotes the area within the convex hull of the detected boundary.
Another limitation placed upon the chosen boundary is a minimum vortex boundary arclength to avoid small-scale local maxima of the IVD field, arising from numerical noise, being misidentified as a vortex. Minimum arclength should be chosen based on the observed or simulated resolution and relates to the length of isocontour that can be realistically derived from the data \citep{Haller_2016}. A boundary will be chosen with convexity deficiency below a certain threshold and arclength above a certain threshold. 
In our analysis, we have placed a further condition on detecting centres using IVD. Consequently, we will only consider maxima in the greatest 1\% of the field to decrease the number of artificial detections. This percentile will be applied to the normalised and the unnormalised methodology. All other aspects follow the method described by \citet{Haller_2016}, where maximum convexity deficiency is set at 0.09 and the arclength criterion is satisfied in that a detection must encompass at least 9 pixels.

\subsubsection{The $\Gamma$ Method}
The $\Gamma$ method is used to find the boundary and centre of a vortex through the combined use of the $\Gamma_1$ and $\Gamma_2$ functions \citep{Graftieaux_2001}. The functions look for solid body rotation within a local region of the flow, outputting values of between -1 and 1, where the sign denotes the orientation of the spin. The discrete formulations for the functions $\Gamma_1$ and $\Gamma_2$, evaluated at a given point $P$, are defined as 
\begin{equation}
    \Gamma_1 (P) = \frac{1}{N} \sum_{\mathcal{X}} \frac{[\boldsymbol{PM} \times \boldsymbol{U}_M] \cdot \boldsymbol{z}}{\Vert \boldsymbol{PM} \Vert \cdot \Vert \boldsymbol{U}_M \Vert}, \label{eq:G1}
\end{equation}

\begin{equation}
    \Gamma_2 (P) = \frac{1}{N} \sum_{\mathcal{X}} \frac{[\boldsymbol{PM} \times (\boldsymbol{U}_M - \tilde{\boldsymbol{U}}_P)] \cdot \boldsymbol{z}}{\Vert \boldsymbol{PM} \Vert \cdot \Vert \boldsymbol{U}_M - \tilde{\boldsymbol{U}}_P \Vert}.\label{eq:G2}
\end{equation}
In the above equations $\boldsymbol{\tilde{U}}_p$ is the local convective velocity which is defined as 
\begin{equation}
    \tilde{\boldsymbol{U}}_P = \frac{1}{N} \sum_\mathcal{X} \boldsymbol{U},
\end{equation}
that is, the spatial average of the velocity in the state space $\mathcal{X}$.
Let $\mathcal{X}$ be a local neighbourhood of the horizontal velocity field containing $N$ discrete sampling points, centred at $P$. For any $M\in\mathcal{X}$, $\boldsymbol{U}_P$ and $\boldsymbol{U}_M$ denote the velocity field evaluated at $P$ and $M$, respectively, and $\boldsymbol{z}$ is the unit normal to the measurement plane.
A high magnitude of both $\Gamma_1$ and $\Gamma_2$ indicates vortical flow dominating that domain.

The function $\Gamma_2$ defines the boundary of a vortex, and the function $\Gamma_1$ defines the centre. If the magnitude of $\Gamma_1$ and $\Gamma_2$ surpass the user-chosen thresholds, a positive vortex detection result will be returned. Lower threshold values (with a range of [0,1]) will increase the number of detections, thus increasing the risk of false positives, whereas higher thresholds will decrease the number of detections whilst increasing the probability of missing detections. Finally, decreasing the $\Gamma_2$ threshold will also increase the area of the vortices detected. For our analysis we employ the thresholds $\Gamma_{1_{thr}} = 0.75$ and $\Gamma_{2_{thr}} = \frac{2}{\pi}.$

In Eqs. (\ref{eq:G1}) and (\ref{eq:G2}), we evaluate $\Gamma_1$ and $\Gamma_2$ at each point $P$ using a local neighbourhood $\mathcal{X}(P)$. The size of $\mathcal{X}$ is a user-defined parameter and directly affects the estimated vortex centre location and the size and shape of detected boundaries. This choice depends on the spatial resolution of the data and the vortex size in the domain. The quantity $\mathcal{X}$ must be large enough to contain the vortex, but not so large that it includes neighbouring vortices, which can bias the detected centre and result in missed detections \cite{Yuan_2023}. If $\mathcal{X}$ is too small, the calculation becomes more sensitive to noise and increases the risk of false positives \cite{Xie_2025}. In this paper, we use an $8 \times 8$ neighbourhood (pixels) to calculate the $\Gamma$ variables.

\section{Results}
We analysed 220 frames spanning $\approx$ 36 minutes over the whole 24 × 24 Mm field of view; our analysis starts $t= 2490$ sec. For each frame and each method, we recorded convexity deficiency, detection counts, vortex area and closed boundaries. We compare detections from the unnormalised (raw) velocity field (AVF) with detections after unit-vector normalisation (the UVF). The number of vortices found over the whole detection process, for each method, can be seen in Table \ref{Table_1}. AVF detection counts vary substantially between methods, with the greatest discrepancy observed for the $\Gamma$ method (5,165 detections) and the $\lambda_2$ criterion (118,037 detections). However, applying the normalisation step greatly reduces the spread in detections between methods. The Q-criterion gives the most detections (25,594), while the $\Gamma$ method again gives the fewest (5,171). The Q and $\lambda_2$ criteria yield much higher AVF detection frequencies than the IVD and $\Gamma$ methods, and both show a large decrease when switching to the UVF (Q: $\approx 55{,}000$ fewer detections; $\lambda_2$: $\approx 100{,}000$ fewer detections). Using the UVF increases detections for the IVD and $\Gamma$ methods, but only modestly (increase of $\approx 3{,}000$ for IVD and $6$ for $\Gamma$). Overall, the UVF yields more consistent total vortex detections across methods.

\begin{table*}[t!]
  \caption{Frequency of detections for each method over the whole time series. }
  \begin{ruledtabular}
  \begin{tabular}{lcccc}
    & Q-criterion  & $\lambda_2$-criterion & IVD & $\Gamma$\\
    \hline
    Total AVF Detections   & 79,966 & 118,037 & 14,729 & 5,165\\
    Total UVF Detections   & 25,594 & 16,814 & 17,600 & 5,171 \\
  \end{tabular}
  \end{ruledtabular}
  \label{Table_1}
\end{table*}

\begin{figure*}[htbp]
  \centering

  \begin{adjustbox}{max width=\textwidth, max totalheight=1\textheight}
    \includegraphics[width=10\linewidth]{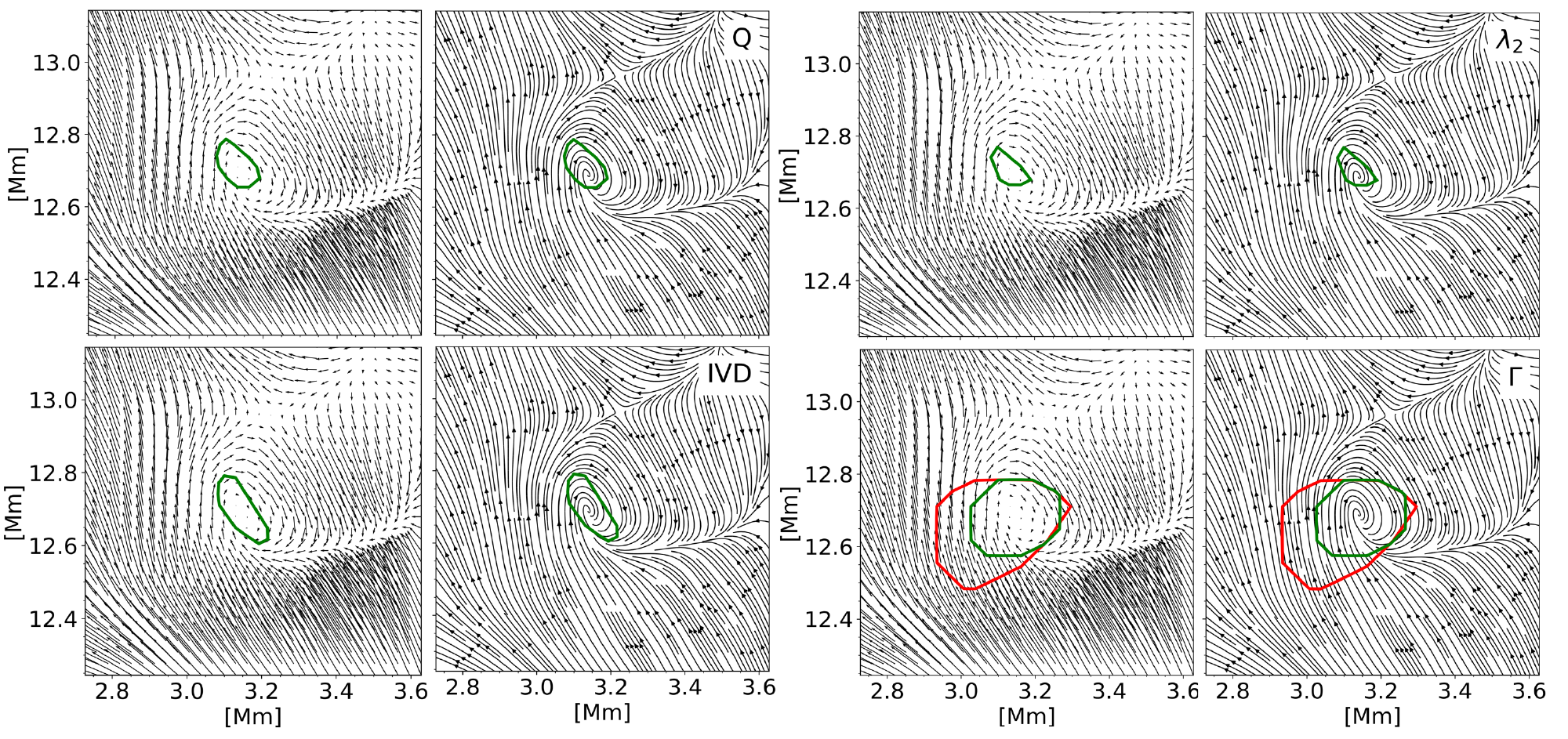}
  \end{adjustbox}

  \caption{Detection results for a structure across all four techniques. The boundary associated with the AVF detection is displayed in red, and the UVF detected boundary in green. If no such boundary is shown, the method did not flag a vortex in this location. The horizontal unnormalised velocity field is shown in quiver plots and streamlines for each method.}
  \label{fig:detection_results}
\end{figure*}

\subsection{Differences in Local Detection Criteria}
Applying the unit-vector preprocessing affects the detection methods in this analysis differently. For example, \(\Gamma_1\) is unchanged by the normalisation, so only \(\Gamma_2\) differs between the AVF and UVF processes. As a result, most vortices detected with the AVF are also recovered with the UVF. However, for non-\(\Gamma\) methods, unit-vector normalisation can substantially change what counts as a vortex locally, potentially producing UVF-only or AVF-only detections.

To establish whether an AVF detection and a UVF detection refer to the same physical structure, we compare their regions and accept a match if at least one of the two overlap criteria is satisfied. In the definitions of these criteria, let \(A\) denote the UVF-detected region and \(B\) the AVF-detected region, and let \(\lvert\cdot\rvert\) denote area. Two vortices, A and B, are considered to detect the same region if
\begin{equation} \label{3.1}
J(A,B) \equiv \frac{\lvert A \cap B \rvert}{\lvert A \cup B \rvert} \ge 0.5,
\end{equation}
which requires at least half of the union area to be shared (Jaccard index /IoU). If Equation (\ref{3.1}) is not satisfied, we require two overlapping vortices to have
\begin{equation}
\mathrm{OC}(A,B) \equiv \frac{\lvert A \cap B \rvert}{\min\{\lvert A \rvert,\lvert B \rvert\}} \ge 0.5,
\end{equation}
where $OC$ denotes the overlapping coefficient, so that a smaller detection largely contained within a larger one is still counted as the same object.

The overlap percentages for the full run are reported in Table (\ref{Table_2}). In Equations (\ref{3.1})-(\ref{eq:nonov-uvf-rate}), we denote the number of UVF detections by  \(n_{\mathrm{UVF}}\), the number of AVF detections by \(n_{\mathrm{AVF}}\) and \(n_{\mathrm{ov}}\) is the number of matched overlaps (per the criteria above). Consequently, we can define
\begin{align} 
\text{AVF detections not matched by UVF}  &= \frac{n_{\mathrm{AVF}}-n_{\mathrm{ov}}}{n_{\mathrm{AVF}}}\times 100\% \label{eq:nonov-avf-rate} \\
\text{AVF detections matched by UVF}      &= \frac{n_{\mathrm{ov}}}{n_{\mathrm{AVF}}}\times 100\% \label{eq:ov-avf-rate} \\
\text{UVF detections not matched by AVF}  &= \frac{n_{\mathrm{UVF}}-n_{\mathrm{ov}}}{n_{\mathrm{UVF}}}\times 100\% \label{eq:nonov-uvf-rate}
\end{align}

\begin{table*}[t!]
  \caption{Summary of local UVF and AVF agreement in detections. }
  \label{tab:final-results}
  \begin{ruledtabular}
  \begin{tabular}{lcccc}
    & Q-criterion  & $\lambda_2$-criterion & IVD & $\Gamma$ \\
    \hline
    AVF detections not matched by UVF    & 84.5\% & 90.0\% & 79.0\% & 0.0\%  \\
    AVF detections matched by UVF        & 15.5\% & 10.0\% & 21.0\%  & 100\%  \\
    UVF detections not matched by AVF    & 51.0\% & 29.7\% & 85.4\% & 0.2\%  \\
  \end{tabular}
  \end{ruledtabular}
  \label{Table_2}
\end{table*}

Table (\ref{Table_2}) highlights the agreement between the UVF and the AVF detections under the $\Gamma$ method. By contrast, agreement is significantly lower for the $\lambda_2$-criterion, the Q-criterion and the IVD method. Unit-vector normalisation as a preprocessing step largely alters the detection outcomes for these three methods. This is expected, as replacing the AVF with the UVF in each method changes the diagnostic values at every grid point. By removing speed information, the UVF reduces the influence of magnitude variations and emphasises the directional (topological) structure of the flow. Switching from the AVF to the UVF, therefore, affects the locations and magnitudes of local maxima and minima, and can shift the threshold values required for a positive vortex detection within a given spatial domain. However, without an independent proxy ground truth, we cannot determine whether these differences represent improvements.

\subsection{The $d$-criterion as a Pseudo Ground-Truth}
The $d$-criterion \citep{Silva_2018} identifies vortex centres using only the local displacement vectors around a candidate point. By examining the displacement vectors immediately north, south, east, and west of this point, we assess whether the flow rotates about it. In solar applications, this idea has been used successfully to identify vortex centres \citep{Silva_2020}, while the associated boundaries can be deduced using other appropriate fields.

Following \citet{Silva_2018}, we compute the $d$-criteria by integrating trajectories over $\Delta t=1\,\mathrm{s}$ while keeping the velocity field fixed at the snapshot time, so that the resultant displacements depend only on the instantaneous velocity field. We also negate the velocity about the central point, let \(D_x\) and \(D_y\) denote the one-second displacement components. The criteria, therefore, can be written as
\begin{equation}
\begin{aligned}
\text{clockwise:}\quad & D_x(i,j+1)>0,\; D_y(i-1,j)>0,\\
                       & D_x(i,j-1)<0,\; D_y(i+1,j)<0,\\[0.5ex]
\text{anticlockwise:}\quad & D_x(i,j+1)<0,\; D_y(i-1,j)<0,\\
                           & D_x(i,j-1)>0,\; D_y(i+1,j)>0.
\end{aligned}
\end{equation}
Here, the first four conditions correspond to a clockwise spinning vortex and the final four to an anticlockwise oriented vortex.

We use the $d$-criterion to validate detections: if a $d$-criterion centre lies within a detected boundary, we treat that detection as a true vortex; detections without a $d$-criterion centre are deemed artificial. This step is necessary because a reliable ground truth is rarely available, making it difficult to assess the accuracy of an automated detector across long time series and large spatial domains. We evaluate the accuracy of detections from the unit-vector field, the actual-velocity field and their overlaps.

Table (\ref{Table_3}) summarises the overall performance for each of the four methods, before and after normalisation. Validation statistics across the 220 time frames of the analysed Bifrost data are reported; we define the metrics used to summarise these $d$-criterion-based validation rates in Equations (\ref{eq:avf-acc})-(\ref{eq:ov-acc}). In the equations below \(\tilde{n}\) denotes the number of validated detections (i.e.\ with a $d$-criterion centre), \(n\) the total number of detections and \(n_{\mathrm{OV}}\) the detections identified by both the AVF and the UVF; subscripts AVF or UVF indicate the presence or absence of our preprocessing step. The validation metrics used are defined as
\begin{align}
\text{AVF Accuracy} &= \frac{\tilde{n}_{\mathrm{AVF}}}{n_{\mathrm{AVF}}} \label{eq:avf-acc}, \\
\text{UVF Accuracy} &= \frac{\tilde{n}_{\mathrm{UVF}}}{n_{\mathrm{UVF}}} \label{eq:uvf-acc}, \\
\text{Overlapping AVF Accuracy} &= \frac{\tilde{n}_{\mathrm{OV}}}{n_{\mathrm{AVF}}}, \label{eq:ov-avf-acc} \\
\text{Overlapping UVF Accuracy} &= \frac{\tilde{n}_{\mathrm{OV}}}{n_{\mathrm{UVF}}}, \label{eq:ov-uvf-acc} \\
\text{Overlapping Accuracy} &= \frac{\tilde{n}_{\mathrm{OV}}}{n_{\mathrm{OV}}}. \label{eq:ov-acc}
\end{align}
Here \(\tilde{n}_{\mathrm{OV}}\) is the number of validated detections from the set of overlapping vortices. 

\begin{table*}[t!]
  \caption{Summary of detection performance, determined by the presence of a $d$-criterion centre.}
  \begin{ruledtabular}
  \begin{tabular}{lcccc}
    & Q-criterion  & $\lambda_2$-criterion & IVD & $\Gamma$ \\
    \hline
    AVF Accuracy    & 21.8\% & 16.3\% & 47.9\% & 98.2\% \\
    UVF Accuracy     & 92.4\% & 98.0\% & 97.0\% & 98.6\% \\
    Overlapping AVF Accuracy     & 14.5\% & 9.8\% & 20.5\% & 98.1\%\\
    Overlapping UVF Accuracy    & 46.5\% & 69.3\% & 14.4\% & 98.5\%  \\
    Overlapping Accuracy    & 94.3\% & 98.4\% & 98.3\% & 98.4\%  \\
  \end{tabular}
  \end{ruledtabular}
  \label{Table_3}
\end{table*}

\begin{figure*}[t!]
  \centering
  \captionsetup{font=small,skip=2pt}
  \begin{adjustbox}{max width=\textwidth, max totalheight=0.43\textheight}
    \includegraphics[width = \linewidth]{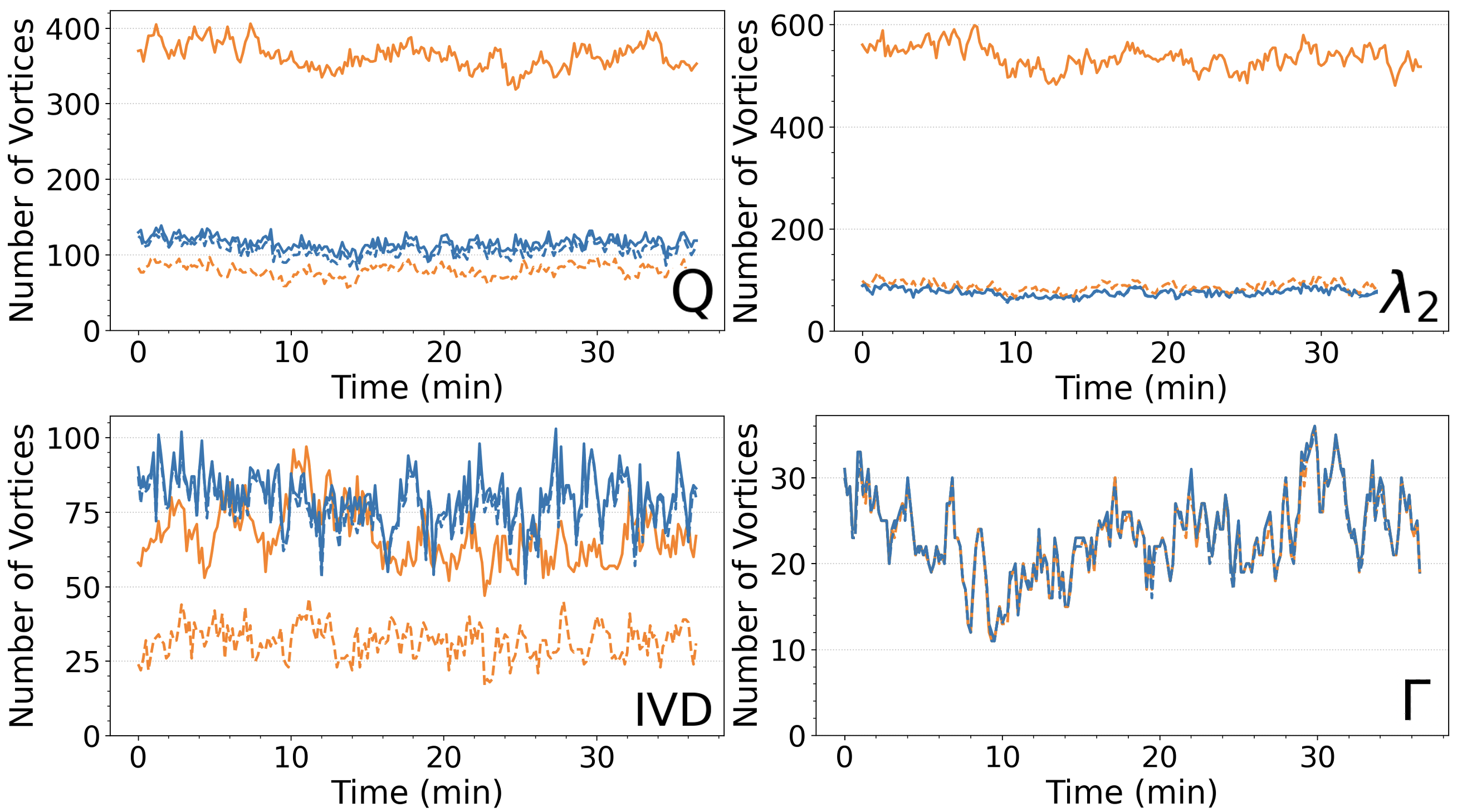}
  \end{adjustbox}
  \caption{The number of reported detections per each time frame for the: AVF detections (orange solid line), validated AVF detections (orange dashed line), UVF detections (blue solid line) and the validated UVF detections (blue dashed line).}
  \label{time_series}
\end{figure*}

Comparing AVF and UVF accuracies shows a significant increase in true detection rates with normalisation for all methods except the $\Gamma$ method, where the true detection rate changes very little between UVF and AVF. Combining these percentages with the counts in Table (\ref{Table_1}) gives the total number of validated detections for the AVF and the UVF. The corresponding time series are shown in Fig. (\ref{time_series}). The validated-overlap statistics quantify how often a true detection from one pipeline is missed by the other. For example, for the $\lambda_2$-criterion, 16.3\% of AVF detections contain a $d$-criterion centre; of these, 9.8\% are detected by both pipelines. This leaves 6.5\% as unique validated AVF detections.
The rates of true and false detections are a key consideration in vortex detection results. Another factor is the frequency of true vortices detected. As shown in Fig. (\ref{time_series}), the number of vortices detected, per time frame, with the use of the $\Gamma$ method is largely unchanged by the normalisation. In the IVD case, the number of detections is changed but not significantly in either direction (more or fewer detections). However, for the $\lambda_2$- and Q-criterion (where the same percentile threshold is applied to the AVF and the UVF methodologies), normalising the velocity field substantially reduces the number of detections. Dashed lines indicate validated detections for the UVF and the AVF. For these two methods, although the AVF yields a higher per-frame count, the number of validated detections remains significantly lower than that count.  

Figure (\ref{time_series}) shows the low per-frame detection counts for the $\Gamma$ method under both the UVF and AVF. Despite this continued underestimation of vortex numbers, nearly all AVF and UVF detections are validated and are thus treated as true vortices under the $d$-criterion ground truth assumption. A similar agreement between total and validated counts is also evident for $\lambda_2$ and IVD under the UVF. For the Q-criterion, however, the UVF yields slightly fewer validated detections than total detections. Using the AVF, the Q-criterion and $\lambda_2$-criterion show a large gap between total and validated detections per frame, indicating the low quality of performance without the normalisation preprocessing step. The IVD method exhibits the same behaviour but to a lesser extent, and still yields improved results when applying the UVF.

\subsection{The Effects of the UVF on Area and Convexity Deficiency}
We assess boundary quality using convexity deficiency (see Equation \ref{eq:G2}), following the commonly used assumption in solar vortex detection that coherent/stable vortex regions are bounded and typically convex \citep{Tziotziou_2023}. In the \(24\times24~\mathrm{Mm}\) domain, the unnormalised velocity field produces a much larger number of detections, many of them false positives - especially for the Q-criterion and the $\lambda_2$-criterion. To focus on true vortices, all convexity deficiency and area results below are restricted to detections validated by an $d$-criterion centre. The validated convexity deficiency distributions are shown in Fig. (\ref{Normalised histograms}) and the validated area distributions in Fig. (\ref{Normalised histograms_1}).
Within the $\Gamma$ method, \(\Gamma_1\) is direction-based and therefore invariant to unit-vector normalisation, so detections are essentially unchanged between the AVF and the UVF. By contrast, \(\Gamma_2\), which defines the boundary, depends on velocity magnitudes; using the UVF methodology, it tends to yield slightly smaller, more convex boundaries (lower convexity deficiency). The $\lambda_2$-criterion shows a similar tendency toward lower convexity deficiency using the UVF modification. In contrast, for IVD and the Q-criterion, using the UVF shifts the convexity deficiency distribution of validated vortices towards higher values, indicating worse convexity than the AVF.

Across all four methods, switching to the UVF generally reduces the detected vortex area and produces boundaries that are more localised about the centre of rotation. The notable exception is the $\Gamma$ method: whether run on the AVF or UVF, it produces vortices that are, on average, substantially larger than those obtained using the other detection methods. The validated area distributions are provided in Fig. (\ref{Normalised histograms_1}).
In practical terms, if the boundary convexity is the primary consideration, the UVF is typically preferable for the $\Gamma$ method and the $\lambda_2$-criterion, whereas the AVF often performs better for IVD and the Q-criterion. Generally, if the user is looking for vortices with a larger detected area, the UVF is likely not the appropriate modification for vortex detection. However, without a ground truth in place for vortical domains, we cannot determine whether the (in general) additional area of an AVF detection encompasses vortical flow.

{
\setlength{\textfloatsep}{8pt plus 2pt minus 2pt}
\setlength{\floatsep}{8pt plus 2pt minus 2pt}
\setlength{\intextsep}{8pt plus 2pt minus 2pt}

\begin{figure*}[t]
  \centering\captionsetup{font=small,skip=2pt}
  \begin{adjustbox}{max width=\textwidth, max totalheight=0.43\textheight}  \includegraphics[width=\linewidth]{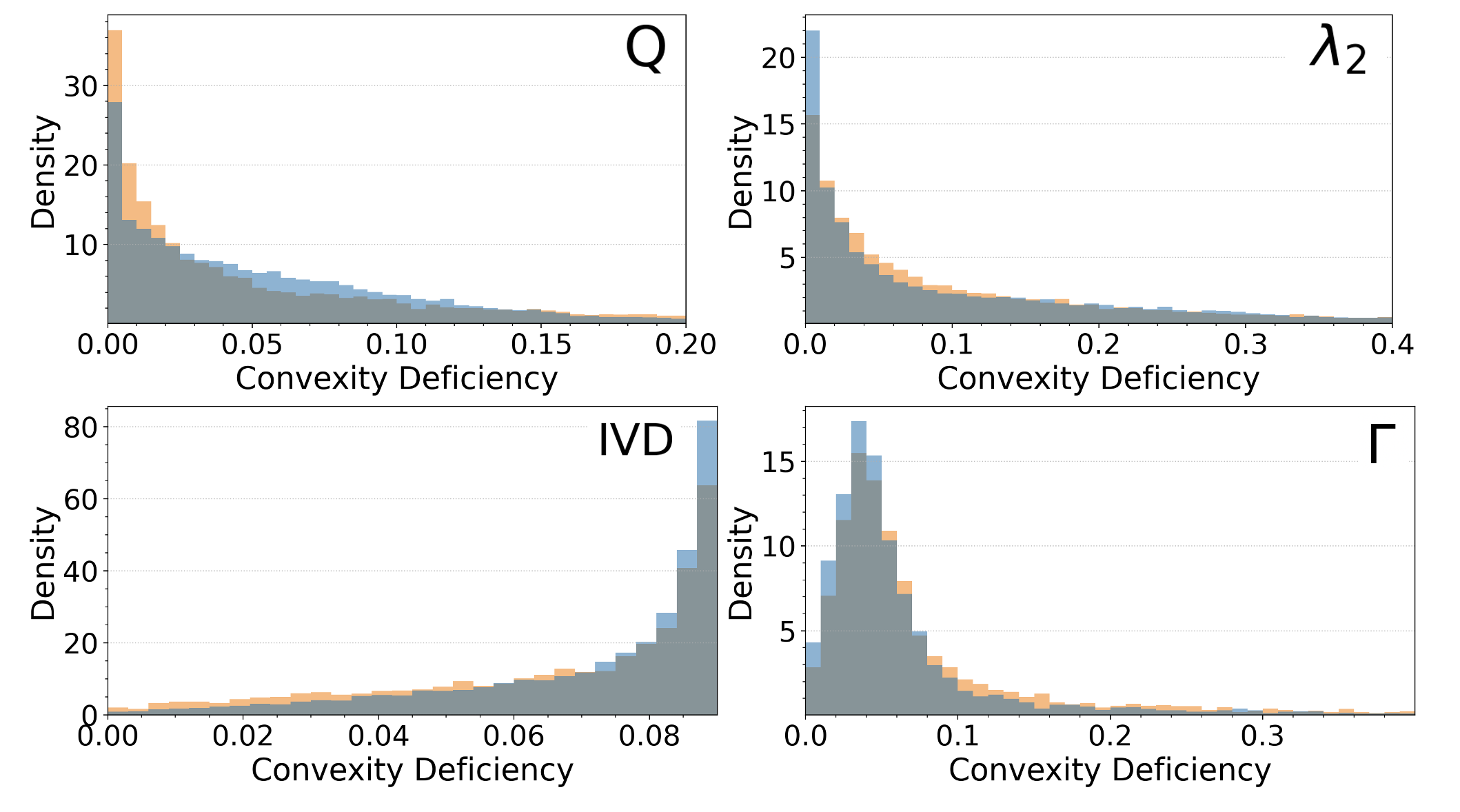}
  \end{adjustbox}
  \caption{Normalised histograms for the convexity deficiency (Eq. \ref{NS}) of UVF and AVF detections validated by a $d$-criterion centre. UVF is shown in blue, while AVF in orange.}
  \label{Normalised histograms}
\end{figure*}

\begin{figure*}[t]
  \centering
  \captionsetup{font=small,skip=2pt}
  \begin{adjustbox}{max width=\textwidth, max totalheight=0.43\textheight}
    \includegraphics[width=\linewidth]{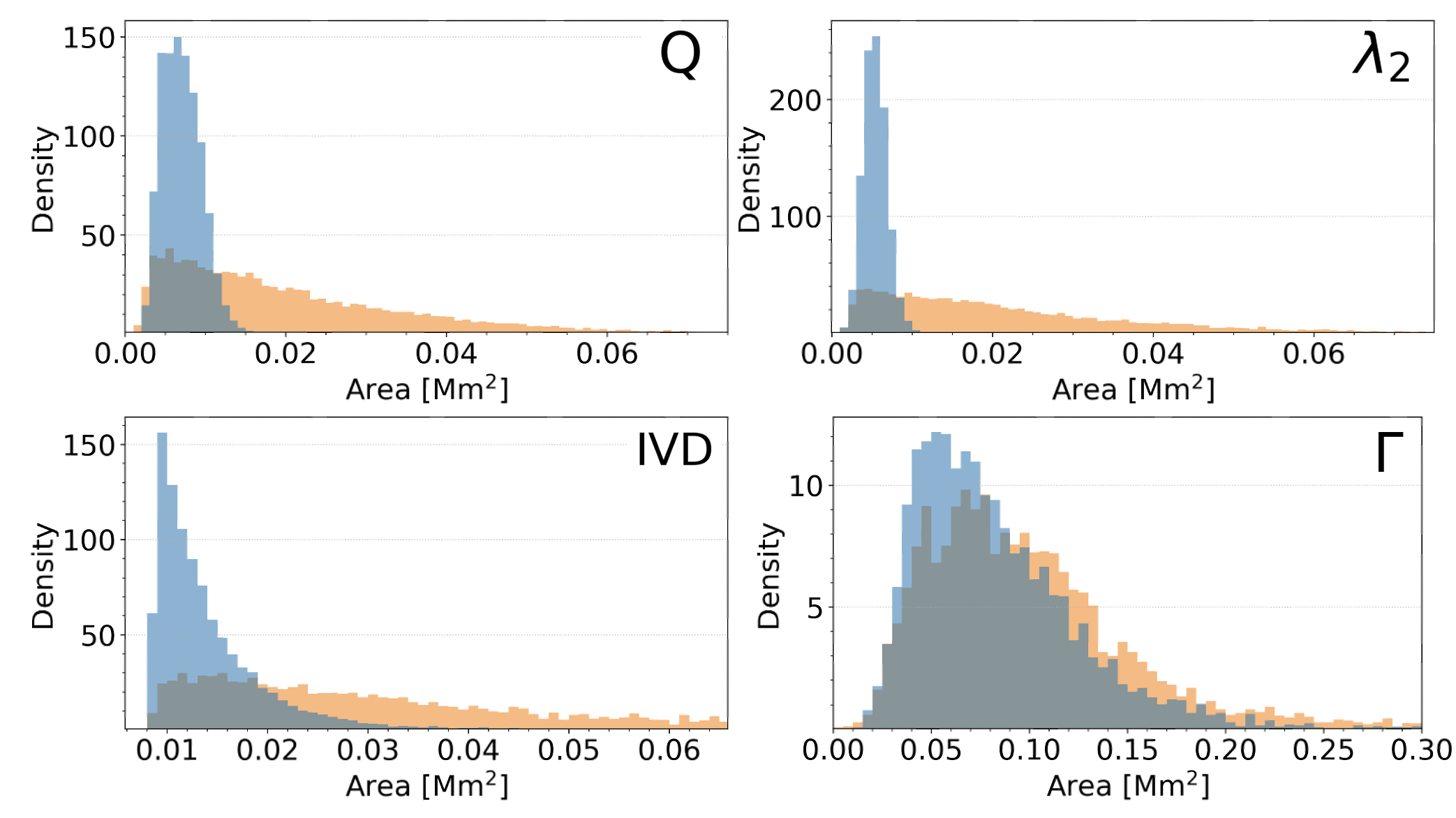}
  \end{adjustbox}
  \caption{Normalised histograms for the area of UVF and AVF detections that have been validated. As before, UVF is shown in blue, AVF in orange.}
  \label{Normalised histograms_1}
\end{figure*}
}

\section{Discussion}

Unit-vector normalisation, used as a preprocessing step for vortex detection, can substantially improve detection outcomes. The changes it induces will differ depending on the method. It is a computationally efficient and easy step to incorporate into any analysis, even when working with large domains over long time series.
We have shown that normalisation changes the shape and location of detected vortices and affects their agreement with the $d$-criterion centres. All methods benefit from the unit-vector velocity field. Although the $\Gamma$ method shows minimal change in validated detection counts, it exhibits better boundary definition, indicated by a shift toward lower convexity deficiency (rounder vortices) when the UVF is applied. This implies that UVF detections have more convex boundaries, consistent with the literature expectation that vortices are approximately convex \citep{Haller_2016, Tziotziou_2023}. The effect is evident in Fig. (\ref{Normalised histograms}) and further illustrated in Fig. (\ref{fig:detection_results}), where the UVF boundary is more tightly focused around the centre of rotation. The $\Gamma$ method, while highly accurate, tends to underestimate the number of vortices. Therefore, switching between the AVF and the UVF has little impact on the overall validity of $\Gamma$ detections, which predominantly identify true vortices.

Noticeable improvements are observed for the other three methods: the Q-criterion, the $\lambda_2$-criterion and the IVD method. For each of these methods, over \(90\%\) of UVF detections are validated as true vortices. We adopt the $d$-criterion as a proxy ground truth: detections containing a $d$-criterion centre are treated as real; those without are treated as false detections. Under this assumption, for the unnormalised versions of the three methods, fewer than \(50\%\) of detections are real. The $d$-criterion evaluates whether detected structures exhibit rotation around a centre, thereby acting as a physical validity check and suppressing false positives arising from shear or noise. Since vortices preferentially form along intergranular lanes, where the flow is dominated by strong shear, vortex detection methods are prone to identifying many spurious structures. When the $d$-criterion is applied to the unnormalised outputs of the three methods, fewer than 50\% of their detections correspond to genuine vortices. Using the overlap criteria presented in Section~3.1 and Table (\ref{Table_3}), we find that for the Q- and $\lambda_2$-criteria only \(\approx 5\%\) of all detections across the 220 time frames are both unique to the AVF and validated (i.e. not also detected by the UVF). For all three methods, when an AVF detection overlaps a UVF detection (per our criteria), it has a $> 90\%$ probability of being validated.

The Q-criterion and the $\lambda_2$-criterion both show a substantial drop in detection count when the velocity field is normalised. 
This is expected, since these methods respond strongly to shear, which often becomes indistinguishable from vorticity in unnormalised fields. Normalising the field to unit magnitude suppresses the shear component, thereby removing many false positives that appear in the AVF detections shown in Fig. (\ref{time_series}).
Even without the $d$-criterion, the sheer number of AVF detections per time frame for the Q- and $\lambda_2$-criteria suggests many are false. Despite these high AVF counts, only \(10\%\) of $\lambda_2$ detections and \(15.5\%\) of Q detections overlap significantly with UVF detections. This indicates that, for the Q and $\lambda_2$ methods, normalising the velocity field before computation substantially improves detection outcomes, as many of the false detections are removed.
Although vortices are inherently local features, normalising the velocity field modifies the local gradient structure in a spatially non-uniform way, since each point is scaled by the inverse of its local speed. This alters the balance between rotation and strain captured by gradient-based criteria. For the Q-criterion, this results in increased sensitivity to small local fluctuations, producing fragmented and less convex (round) vortex boundaries after normalisation. In contrast, the $\lambda_2$-criterion is based on the coherent eigenstructure of $\boldsymbol{S}^2 + \boldsymbol{\Omega}^2$ and is, therefore, less affected by such fluctuations, resulting in more compact and round vortex regions.

Using the UVF generally reduces the vortex areas returned by each method. If we estimate the energetic contribution of vortical motions by integrating only over these detected regions, the result will be underestimated because part of the rotational flow is excluded. This is significant because the highest flow speeds in a vortex typically occur away from the centre, toward the outer part of the rotating structure, as described in classical hydrodynamical models such as those of Lamb-Oseen and Burgers \citep{Marshall2025} and observed in vortical plasma flows \citep{Silva_2020}. An underestimated boundary can, therefore, exclude a disproportionate share of the kinetic-energy contribution. Across a system of vortices, both smaller detected areas and missed detections will lead to an underestimation of the role of vortical motion in the overall solar atmospheric dynamics and energetics. 
We should note that this potential underestimation is primarily inferred from visual comparison with streamlines. Moreover, larger detected regions do not necessarily provide a more accurate estimate of vortex extent. The study by \cite{Rempel_2016} showed that even slightly enlarged boundaries may include regions that are not co-rotating with the vortex and therefore do not represent true vortical motion. Likewise, regarding energy estimates, \cite{Silva_2024b} demonstrated that even when the detected boundary appears underestimated, the largest Poynting fluxes are still located just within the boundary region. Thus, although reduced UVF areas could suggest an energetic underestimation, previous studies indicate that larger boundaries may not yield more accurate estimates, either in terms of vortex extent \citep{Rempel_2017} or energetic contribution \citep{Silva_2024b}. In the absence of a definitive ground truth, it remains unclear which boundary definition provides the most physically meaningful estimate.
This caveat is particularly relevant for the $\Gamma$ method (and other cases where detection frequencies are low), where under-detection is combined with reduced detected areas under the UVF. Conversely, AVF-based detections with the $\lambda_2$- and Q-criterion can overestimate the number of vortices in the system by over-detecting features that fail validation against the $d$-criterion ground truth; in this case, the inferred contribution of vortical motion to the flow may be artificially inflated.

\section{Conclusion}

This study demonstrates that vortex detection in solar plasma flows can be substantially improved by normalising the velocity field, which yields more physically realistic vortex regions. When combined with the $d$-criterion, the use of our method significantly increases the number of reliable detections. This refinement is particularly relevant for studies of the solar atmosphere, where vortices may play an important role in energy transport. Our results show that the $\Gamma$ method considerably underestimates vortex occurrence. Since the $\Gamma$ method has been the basis for most observational analyses, existing studies likely capture only a subset of the vortical flows present. This is in accordance with earlier results by \cite{Silva_2018} for vortical flows in supergranular junctions. Our analysis shows that the $Q$-criterion, $\lambda_2$-criterion, and IVD detect roughly four to five times more vortices than the $\Gamma$ method. Using the $\Gamma$ method, \citet{Giagkiozis_2018} estimated that vortices cover 2.8\% of the photospheric surface; according to our estimates for $\Gamma$ method limitations, the true value may lie between 11.2\% and 14\% (based on the results and methodology laid out in \cite{Giagkiozis_2018}). As vortices are key drivers of energy transport \citep{Yadav_2020, Yadav_2021, Kuniyoshi_2023, Silva_2024a, Silva_2024b, Kuniyoshi_2025}, reliably identifying these coherent structures in both simulations and observations is therefore essential, and the method presented here enables this.

With ongoing advances in high-resolution simulations and observations, accurate vortex detection is more important than ever. Our results show that this can be particularly challenging in high-shear and turbulent environments, where commonly used methods struggle to provide reliable detections. We conclude that normalising the data as a preprocessing step improves vortex detection across all four examined and compared methods. Without this preprocessing, three of these four methods (Q, $\lambda_2$, and IVD) produce inaccurate or unreliable results. Although all detections made by the $\Gamma$ method are valid, it nevertheless fails to identify a large fraction of the swirling flows present in the data. Such limitations are highly problematic given the unprecedented detail now available from state-of-the-art instrumentation and simulations.

Future work will extend the unit-vector normalisation to additional detecting techniques, including the swirling-strength criterion \citep{ZHOU_1999}, LAVD \citep{Haller_2016}, the advanced $\Gamma$ method \citep{Yuan_2023}, and the vorticity-strength criterion \citep{Kato_2017}, to determine whether normalisation provides universal improvements by minimising asymmetries in vortex flows. Overall, the normalisation procedure presented here is a minimal yet powerful modification that enables robust vortex detection in both simulated and observational data, and should be regarded as a standard step in analyses requiring accurate identification of vortical structures.

Although our study has focused on solar applications of the methodology, the approach is not limited to solar physics. Any study requiring automated vortex detection in complex hydrodynamic environments can benefit from this normalisation step. Vortex identification is a central task in fluid dynamics, from canonical laboratory turbulence to engineering flows \citep{Jeong_Hussain_1995, CHAKRABORTY_2005, Adrian_2007}. In particular, the widely used Q- and $\lambda_2$-criteria are highly sensitive to shear-induced artefacts in regions of strong velocity gradients. This is especially problematic in flows containing numerous antisymmetric vortices, which are typical of many turbulent environments, where the unit-vector field preprocessing step is essential to suppress such artefacts and to isolate physically meaningful coherent structures. Consequently, the normalisation strategy developed here has potential utility across a broad range of hydrodynamic applications in which robust, automated vortex detection is required. Moreover, the combination of normalisation and the $d$-criterion offers a simple, physically motivated procedure that is straightforward to implement and does not depend on case-specific heuristics. This stands in contrast to recent computer-vision and machine-learning approaches \citep[e.g.][]{Abolholl_2023}, which can reduce false classifications but require substantial training data, computational resources, and problem-specific tuning. By comparison, the present method provides a lightweight, physics-based alternative suitable for high-resolution simulations and observational datasets alike.

	\section*{Acknowledgments}
			L.M. is grateful for the STFC studentship and the continued support from her collaborators and supervisors. V.F., G.V. and S.S.A.S. are grateful to the Science and Technology Facilities Council (STFC) grants ST/V000977/1, ST/Y001532/1. V.F., S.S.A.S., I.B. and G.V. are grateful to The Royal Society, International Exchanges Scheme, collaboration with Greece (IES/R1/221095) and The Royal Society, IEC/R3/233017 - International Exchanges 2023 Cost Share (NSTC), collaboration with Taiwan. L.M., S.S.A.S. and V.F. would like to thank the International Space Science Institute (ISSI) in Bern, Switzerland, for the hospitality provided to the members of the team on `Opening new avenues in identifying coherent structures and transport barriers in the magnetised solar plasma’. L.M. wishes to thank Ian McClure for all his support.

\begin{figure*}[t!]

  \centering
  \captionsetup{font=small,skip=2pt}
  \begin{adjustbox}{max width=\textwidth, max totalheight=1\textheight}
    \includegraphics[width=10\linewidth]{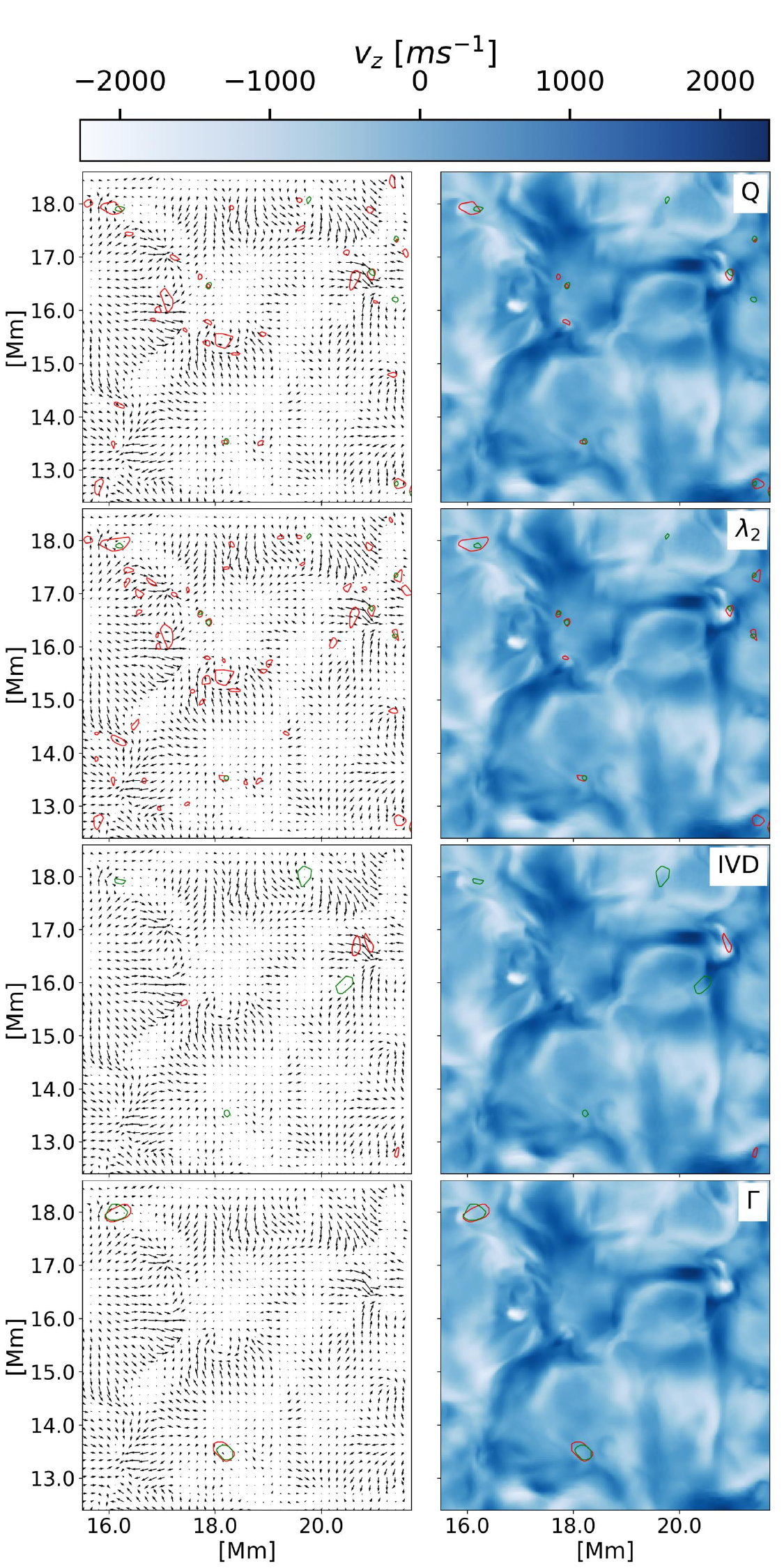}
  \end{adjustbox}
  \caption{Here we display a subset of the domain, in which we visualise how successful each vortex detection method is when using the $d$-criterion centres as a ground truth. The left column shows all vortex detection results from each detection method; the UVF results are shown in green, and the AVF results are plotted in red. The remaining detections in the right-hand side column show only those that have been validated by a $d$-criterion centre. These validated vortices are plotted against the $z$-component of velocity. }
\label{fig:6}
\end{figure*}

\begin{figure*}[t!]

  \centering
  \captionsetup{font=small,skip=2pt}
  \begin{adjustbox}{max width=\textwidth, max totalheight=1\textheight}
    \includegraphics[width=10\linewidth]{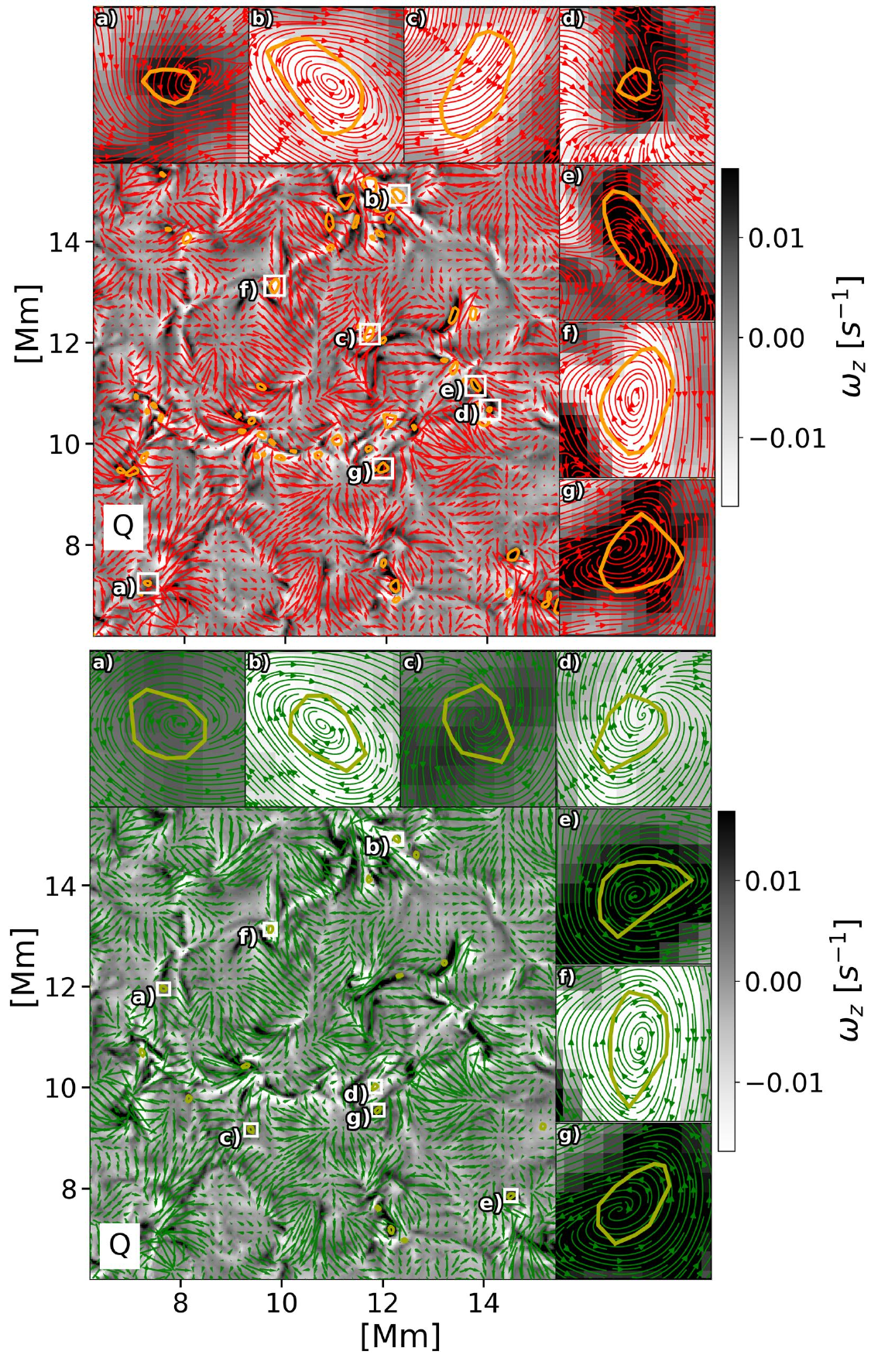}
  \end{adjustbox}
  \caption{A subdomain at $t = 0$ (index number 249), where we show the detection results for the Q-criterion on a quiver plot of the velocity field. The background shows the $z$-component of vorticity. The top panel (red) shows the AVF detections, and the bottom panel (green) displays the UVF detection results. In the tiles surrounding each panel, we show a zoomed-in picture of the convex hulls for a selection of the detections with streamlines tracing the unnormalised horizontal velocity field. }
\label{fig:7}
\end{figure*}

\begin{figure*}[t!]

  \centering
  \captionsetup{font=small,skip=2pt}
  \begin{adjustbox}{max width=\textwidth, max totalheight=1\textheight}
    \includegraphics[width=10\linewidth]{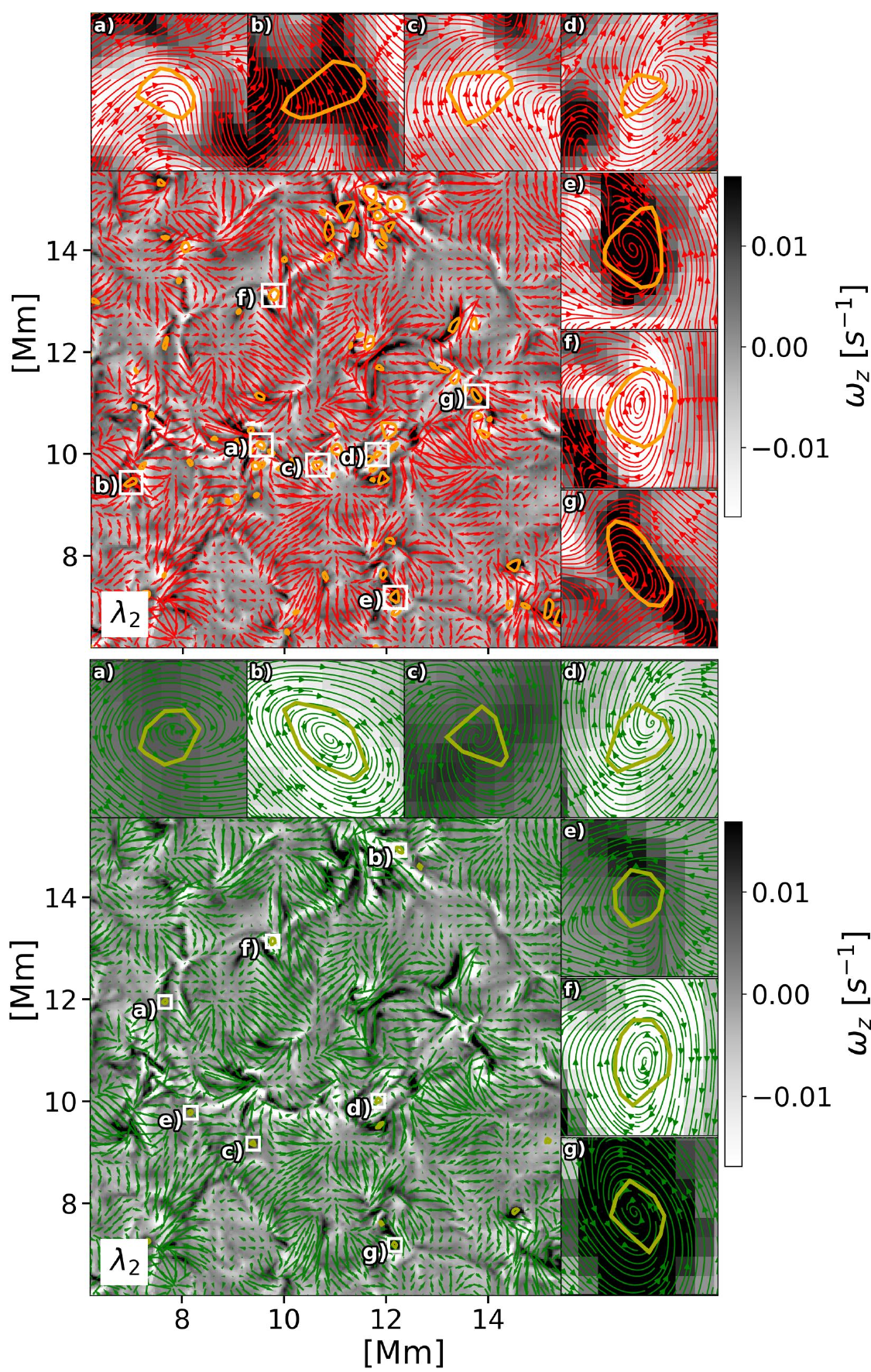}
  \end{adjustbox}
  \caption{The same domain as in Fig. (\ref{fig:7}), with detection resulting from the use of the $\lambda_2$-criterion.} 

\label{fig:8}
\end{figure*}

\begin{figure*}[t!]

  \centering
  \captionsetup{font=small,skip=2pt}
  \begin{adjustbox}{max width=\textwidth, max totalheight=1\textheight}
    \includegraphics[width=10\linewidth]{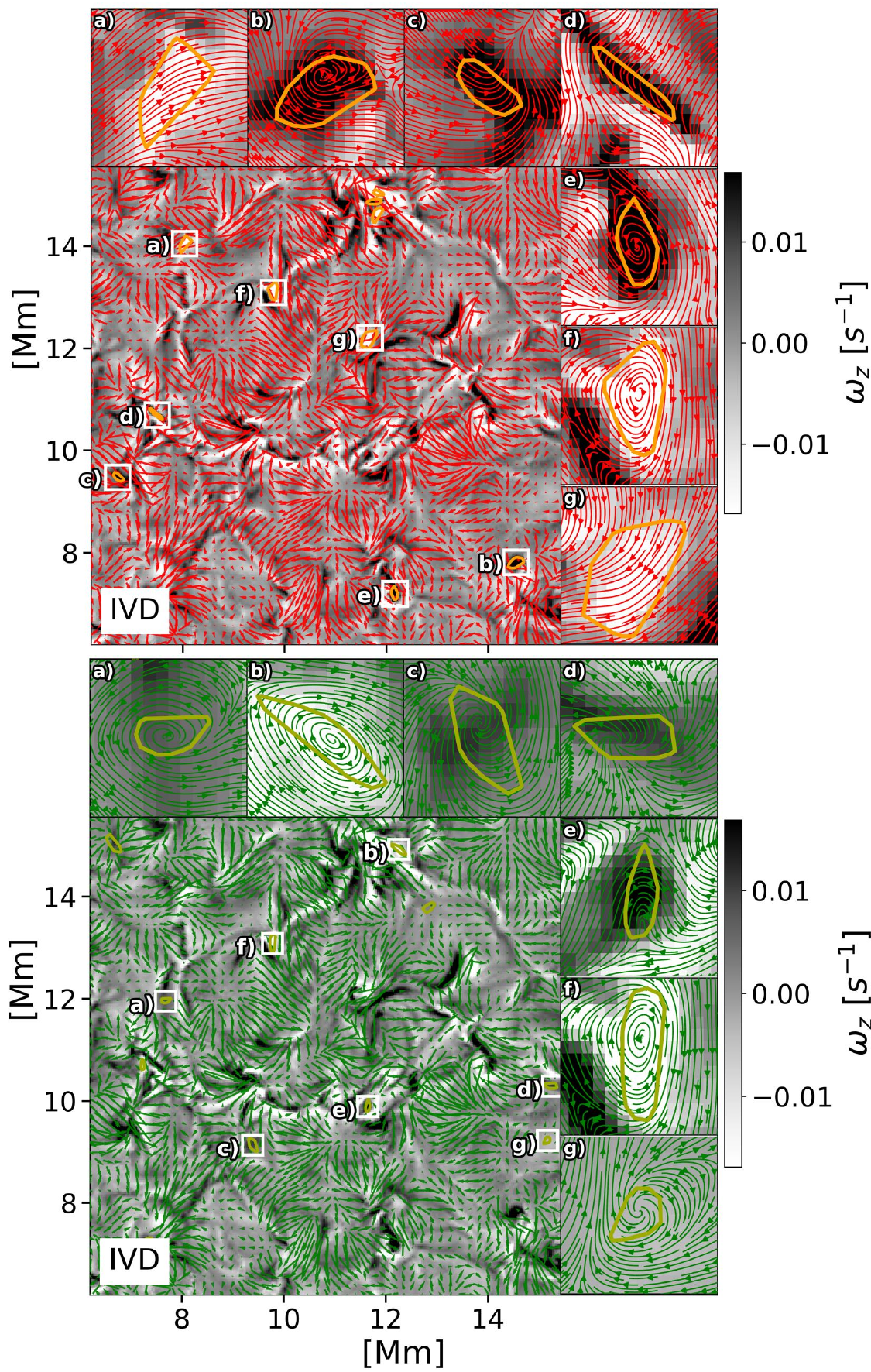}
  \end{adjustbox}
  \caption{The same domain as in Fig. (\ref{fig:7}), with detection resulting from the use of the IVD method.}
\label{fig:9}
\end{figure*}

\begin{figure*}[t!]

  \centering
  \captionsetup{font=small,skip=2pt}
  \begin{adjustbox}{max width=\textwidth, max totalheight=1\textheight}
    \includegraphics[width=10\linewidth]{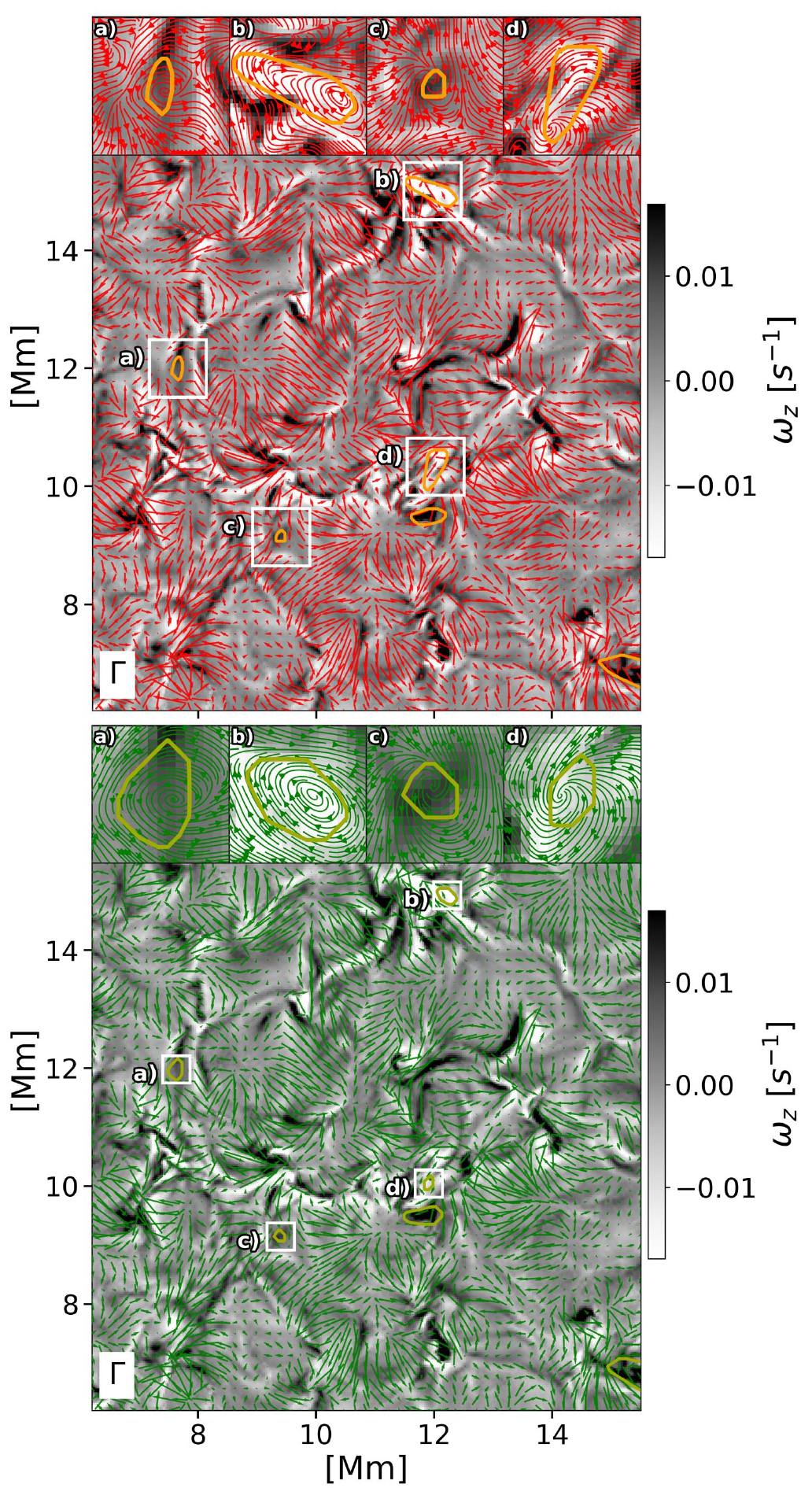}
  \end{adjustbox}
  \caption{The same domain as in Fig. (\ref{fig:7}), with detection resulting from the use of the $\Gamma$ method.}
\label{fig:10}
\end{figure*}

\appendix
In this appendix, we have included larger-scale visualisations of vortex detection results from the Bifrost data. These are provided as validation of the preprocessing normalisation step in each vortex detection methodology. In each figure, a subdomain of the entire data set is presented, and vortex detection results from different detection methods are shown, using both the UVF and the AVF. In Figures (\ref{fig:7})--(\ref{fig:10}) we present results for the Q-criterion, the $\lambda_2$-criterion, the IVD method and the $\Gamma$ method, respectively compared to streamline plots of the velocity field. Figure (\ref{fig:6}) compares all techniques against the $d$-criterion ground truth. 

\bibliography{mybibliography}{}
\bibliographystyle{aasjournal}

\end{document}